\documentclass{article}
\usepackage{ijcai25}

\usepackage{times}
\usepackage{soul}
\usepackage{url}
\usepackage[hidelinks]{hyperref}
\usepackage[utf8]{inputenc}
\usepackage[small]{caption}
\usepackage{graphicx}
\usepackage{amsmath}
\usepackage{amsthm}
\usepackage{booktabs}
\usepackage{algorithm}
\usepackage{algorithmic}
\usepackage[switch]{lineno}

\usepackage{amssymb}
\usepackage{multirow}
\usepackage{graphicx}
\usepackage{appendix}
\usepackage{xcolor}
\usepackage{appendix}

\title{GCTAM: Global and Contextual Truncated Affinity Combined Maximization Model For Unsupervised Graph Anomaly Detection}

\author{
Xiong Zhang${{\dagger}}$\and
Hong Peng${{\dagger}}$\and
Zhenli He \and
Cheng Xie$^*$\and
Xin Jin \And
Hua Jiang \\
\affiliations
School of Software, Yunnan University,  Kunming,  China\\
\emails
zhangxiong@stu.ynu.edu.cn,
\{software\_ph, hezl, xiecheng, xinjin, huajiang\}@ynu.edu.cn,
}

\begin{document}
\maketitle
\begin{abstract}
Anomalies often occur in real-world information networks/graphs, such as malevolent users, malicious comments,  banned users, and fake news in social graphs. 
The latest graph anomaly detection methods use a novel mechanism called truncated affinity maximization (TAM) to detect anomaly nodes without using any label information and achieve impressive results.
TAM maximizes the affinities among the normal nodes while truncating the affinities of the anomalous nodes to identify the anomalies.
However, existing TAM-based methods truncate suspicious nodes according to a rigid threshold that ignores the specificity and high-order affinities of different nodes.
This inevitably causes inefficient truncations from both normal and anomalous nodes, limiting the effectiveness of anomaly detection. 
To this end, this paper proposes a novel truncation model combining contextual and global affinity to truncate the anomalous nodes.
The core idea of the work is to use contextual truncation to decrease the affinity of anomalous nodes, while global truncation increases the affinity of normal nodes.
Extensive experiments on massive real-world datasets show that our method surpasses peer methods in most graph anomaly detection tasks.
In highlights, compared with previous state-of-the-art methods, the proposed method has +15\% $\sim$ +20\% improvements in two famous real-world datasets, Amazon and YelpChi. 
Notably, our method works well in large datasets, Amazin-all and YelpChi-all, and achieves the best results, while most previous models cannot complete the tasks.  
\end{abstract}

\section{Introduction}
\begin{figure}
    \centering
    \includegraphics[width=0.9\linewidth]{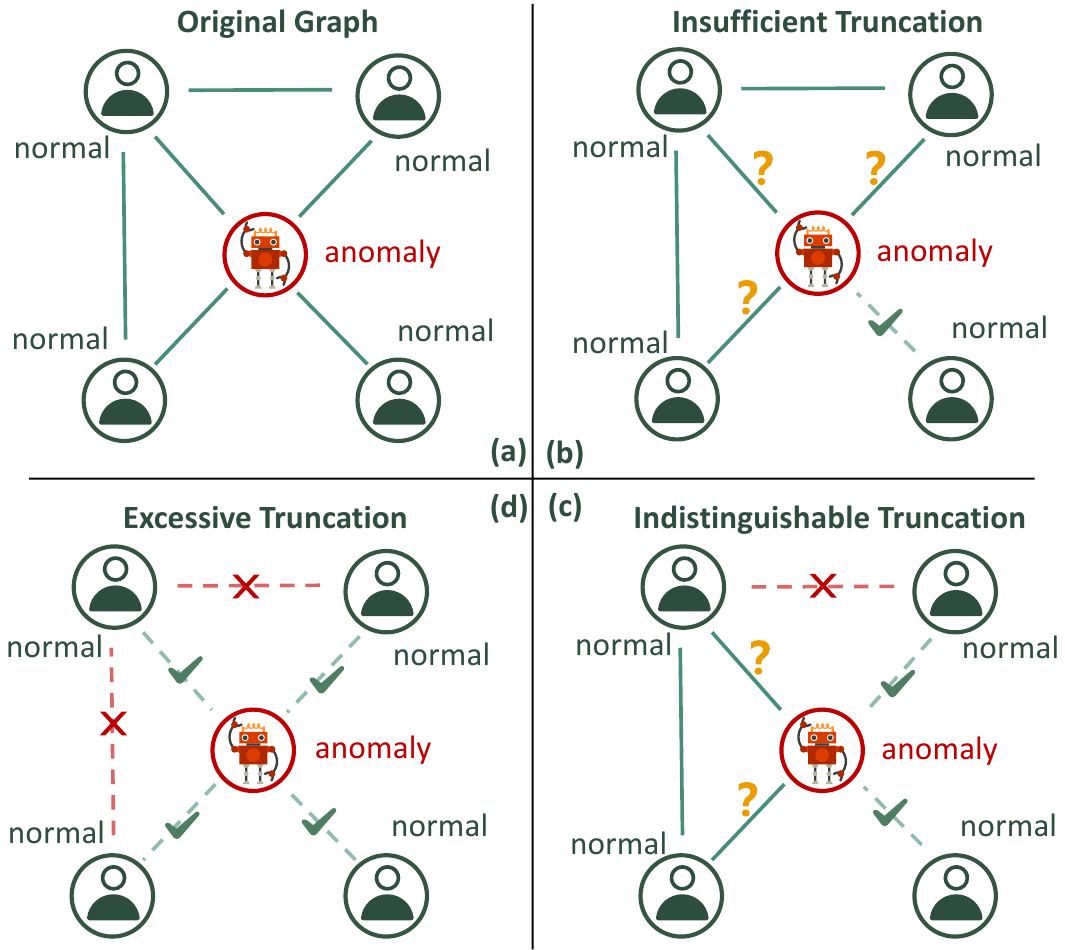}
    \caption{The motivation of the work. (a) original graph with an anomaly. (b) Insufficient Truncation; (c) Indistinguishable Truncation; (d) Excessive Truncation.}
    \label{fig:fig1}
\end{figure}

Graph anomaly detection (GAD) aims to identify patterns that significantly deviate from the majority within a graph. Depending on the detection task, it can be categorized into three levels: node-level~(Fig.\ref{fig:fig1}(a)), edge-level, and subgraph-level. With the widespread availability of information through anonymous accounts on commercial websites and telecommunications networks, these platforms have become prime targets for fraudsters and attackers seeking to spread misinformation, cause disruptions, and engage in malicious activities~\cite{pourhabibi2020fraud}. Due to its broad practical applications, node-level graph anomaly detection has garnered significant research interest~\cite{pourhabibi2020fraud,twbot22_feng2022twibot,tang2022rethinking,dae_fan2020anomalydae}, particularly in areas such as financial transaction networks and social networks. For instance, detecting anomalous accounts is crucial as they may have fraudulent transactions~\cite{tang2022rethinking}. Likewise, social networks often contain anomalous nodes, such as social bots that disseminate rumors and false information~\cite{twbot22_feng2022twibot}. Therefore, effective node-level graph anomaly detection plays a crucial role in enhancing the security and reliability of complex networks by mitigating fraudulent activities and ensuring data integrity. For convenience, all references to graph anomaly detection in the following article will specifically refer to node-level GAD.


Graph neural networks (GNNs) have been widely used in graph anomaly detection (GAD) by leveraging their ability to capture complex graph structures. Due to the frequent updates and changes in real-world networks, it is generally difficult to provide complete label information for GNN models. As a result, traditional supervised learning algorithms are not well-suited for graph anomaly detection tasks. Existing self-supervised GNN-based approaches for GAD can be categorized into reconstruction-based and contrastive-based self-supervised learning methods, such as DOMINANT~\cite{dominant_ding2019deep} and ComGA~\cite{ComGA_luo2022comga}, which detect anomalies by reconstructing adjacency and attribute matrices and identifying nodes with high reconstruction errors, assuming that anomalous nodes deviate significantly from normal patterns. On the other hand, self-supervised learning approaches, including Cola\cite{Cola_liu2021anomaly}, utilize contrastive and generative learning objectives to extract meaningful node representations and identify irregularities based on relational inconsistencies. While reconstruction-based methods minimize the reconstruction error, resulting in numerous cases in which non-trivial nodes are misidentified, contrastive-based learning methods randomly generate the augmented graphs, which become unstable and do not achieve optimal performance in real-world data environments.
In recent years, the truncated affinity maximization (TAM) method~\cite{TAM_qiao2024truncated}, based on the homogeneity assumption, has achieved state-of-the-art results in anomaly detection. This approach learns node representations for anomaly identification by maximizing the truncation affinity between a node and its neighbors while truncating the affinities of the anomalous nodes to identify the anomalies.

Existing TAM methods typically truncate suspicious nodes based on a rigid or predefined threshold, such as one computed using Euclidean distance. However, this approach overlooks the specificity and high-order affinities of different nodes, leading to suboptimal truncation performance. For instance, as illustrated in Fig.\ref{fig:fig1}(b), an insufficient truncation distance threshold may fail to remove all edges between normal and anomalous nodes, resulting in incomplete truncation. Moreover, as shown in Fig.\ref{fig:fig1}(c), an indistinguishable truncation threshold not only fails to effectively separate anomalous edges but may also mistakenly truncate normal edges between normal nodes, further complicating the detection process. Conversely, as depicted in Fig.~\ref{fig:fig1}(d), an excessively high truncation threshold leads to over-truncation, which removes most of the edges of nodes. This, in turn, results in the affinity of the node being calculated as zero, making it impossible to detect anomalies. In summary, the use of rigid truncation thresholds in existing methods significantly hampers the accurate calculation of node affinities, ultimately reducing the effectiveness of anomaly detection.

To this end, we propose GCTAM, a global and contextual truncation affinity combined maximization model for unsupervised graph anomaly detection. Specifically, we introduce a contextual affinity truncation module that truncates the edges between normal and anomalous nodes, effectively reducing anomaly affinity while preserving the affinity of normal node relationships, which enhances the discriminative capability of our framework, thereby ensuring robust performance in downstream anomaly detection tasks.  Secondly, we design a global affinity truncation module based on the homogeneity assumption. This model constructs a global affinity truncation graph that enriches edge connectivity between normal nodes, aiming to increase the affinity of nodes. Doing so further enhances the performance of the contextual truncation affinity module. Finally, we introduce shared parameter graph convolution networks (GCNs) to integrate node representations from both the contextual and global affinity truncation graphs, and the unified representation is better equipped to distinguish the local affinities of normal nodes from those of anomalous nodes.
Through extensive experimental comparisons on seven real-world GAD datasets, empirical results demonstrate that GCTAM outperforms eight competing models. Notably, on more challenging datasets, GCTAM achieves +15\% $\sim$ +20\% improvements in AUROC and AUPRC compared with the best-performing competitor. The code can be found at \url{https://github.com/kgccc/GCTAM}.

In summary, our contributions are as follows:
\begin{itemize}
    \item Based on the homogeneity assumption, we introduce a contextual affinity truncation model to address the challenges of insufficient, indistinguishable, and excessive truncation.
    
    \item We further present a global affinity truncation model that improves node affinities within the global affinity truncation graph, facilitating the contextual affinity truncation model to learn more discriminative node representations for effective anomalous node detection.

    \item We propose the model GCTAM that effectively integrates the contextual affinity truncation (CAT) and global affinity truncation (GAT) modules, enhancing the overall performance through their complementary strengths. Empirical results on seven real-world GAD datasets demonstrate that our GCTAM model significantly outperforms eight competing models.
\end{itemize}

\section{Related Work}
In this section, we briefly introduce reconstruction-based self-supervised learning, contrastive-based self-supervised learning, and affinity-based unsupervised learning methods.  

\subsection{Reconstruction-based Self-Supervised Learning}
Reconstruction-based self-supervised methods usually use a graph auto-encoder (GAE) focusing on learning a node representation of a GAD by minimizing errors in reconstructing node attributes and graph structures. For example, DOMINANT\cite{dominant_ding2019deep} pioneered the first generative GAD method to learn node embeddings by minimizing the reconstruction error of attribute and adjacency matrices while exploiting attribute and topological features. AnomalyDAE\cite{dae_fan2020anomalydae} decouples attribute and structural encoders to efficiently model interactions. ComGA \cite{ComGA_luo2022comga} incorporates community detection. ANOMALOUS \cite{ANOMALOUS_peng2018anomalous} jointly considered CUR decomposition and residual analysis for anomaly detection in attribute networks. These methods ignore the anomaly-discriminative property of abnormal nodes in anomaly monitoring, thus limiting their anomaly detection capabilities.

\begin{figure*}[ht]
    \centering
    \includegraphics[width=1\linewidth]{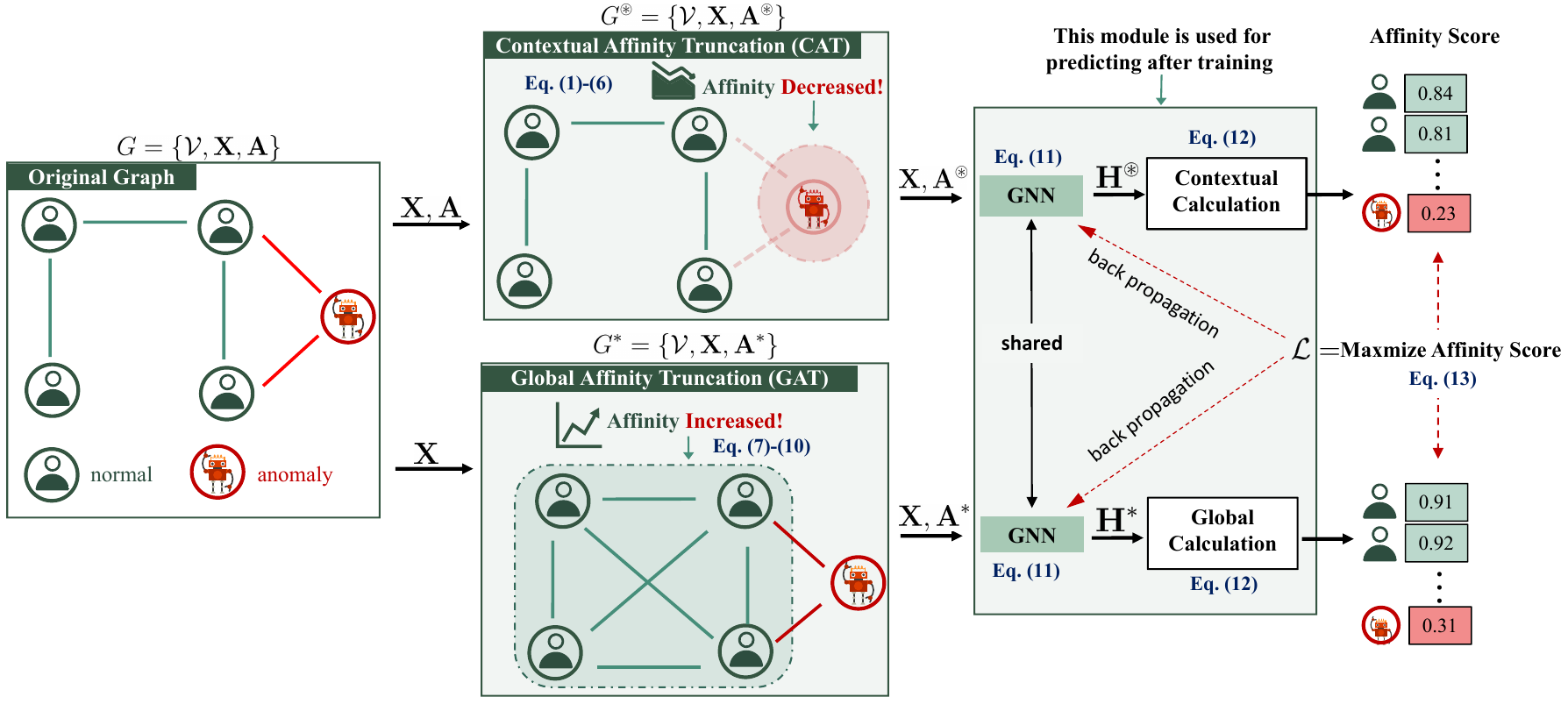}
    \caption{An overview of the GCTAM framework. The contextual truncation graph $G^{\circledast}$ truncates the edges between normal and anomaly nodes to 
    decrease the affinity of anomaly nodes. Conversely, the Global Truncation Graph $G^{*}$ reinforces edges between normal nodes to increase their affinity.}
    \label{fig: frame_work}
\end{figure*}

\subsection{Contrastive-based Self-Supervised Learning}
For contrastive learning-based methods, CoLA \cite{Cola_liu2021anomaly} addresses challenges with a contrastive learning paradigm, using a discriminator to detect inconsistencies between the target node and the neighbor subgraph embeddings. ANEMONE \cite{ANEMONE_zheng2021generative} introduces a patch-level contrastive task for multi-scale anomaly detection. GRADATE \cite{duan2023graph} improves the framework through graph augmentation and multi-view contrast. SL-GAD \cite{zheng2021generative} combines attribute reconstruction and node-subgraph contrast, while Sub-CR \cite{zhang2022reconstruction} uses a masked autoencoder and graph diffusion to fuse attributes with local and global topological information. 
Although these methods build classification models on the relationship between nodes and context subgraphs, this group of methods is not designed for the task of anomaly detection, and their performance depends largely on the relationship between pre-tasks and anomaly detection.

\subsection{Affinity-based Unsupervised Learning}
Apart from the two main categories, an anomaly detection method based on the homogeneity assumption, TAM \cite{TAM_qiao2024truncated}, has achieved the best results by learning the node representations for our anomaly measurements by maximizing the local affinity between a node and its neighbors. Although TAM achieves the best results, it ignores the specificity of different nodes by truncating suspicious nodes based on the rigid Euclidean distance threshold. 

\section{Method}




This section presents GCTAM, a model that integrates contextual and global truncated affinity maximization. As shown in Fig.~\ref{fig: frame_work}, it consists of two key modules: contextual affinity truncation (CAT) and global affinity truncation (GAT). CAT iteratively removes anomalous edges to reduce anomaly affinity while retaining normal ones. GAT constructs a global affinity graph to enhance normal node affinities, further reinforcing CAT’s effectiveness.


\subsection{Preliminaries}

We tackle unsupervised anomaly detection on an attributed graph. 
Supposing that $G=(\mathcal{V}, \mathbf{A},  \mathbf{X})$ is an attributed graph with $N$ nodes, where $ \mathcal{V} = \left\{v_1, \cdots, v_N\right\}$ denotes its node set. 
Then, we denote the matrix $\mathbf{X} \in \mathbb{R}^{N \times d}$ as the attributes' feature of all nodes where $\mathbf{x}_i \in \mathbb{R}^d$ is the $d$-dimensional attribute vector of node $v_i$.
And $\mathbf{A} \in\{0,1\}^{N \times N}$ is the adjacency matrix of graph $G$ with ${{\bf{A}}_{ij} = 1}$ if $v_i$ and $v_j$ are connected. The proposed model aims to learn an affinity score, $\mathcal{AS}(\cdot)$, such that $\mathcal{AS}(v) \gg \mathcal{AS}(v^{\prime})$ for any $v\in \mathcal{V}_{n}, v^{\prime} \in \mathcal{V}_{a}$, where ${{\mathcal V}_n}$ and ${\mathcal V}_{a}$ denotes the set of normal and abnormal nodes, respectively. According to the nature of GAD, it is typically assumed that $\left| {\mathcal V}_n \right| \gg \left| {\mathcal V}_{a} \right|$.

\subsection{CAT: Contextual Affinity Truncation}
In this paper, we propose a module, CAT, designed to preserve edges between normal nodes while effectively truncating edges connecting normal and anomalous nodes based on contextual affinity among their neighbors. 
By avoiding reliance on a fixed Euclidean distance, CAT reduces the risk of incorrect edge truncation. This approach keeps the affinity among normal nodes while effectively decreasing the affinity of anomalous nodes.
In detail, it consists of two important processes: (1) selecting the normal edge to be preserved based on each node's cosine similarity with its neighbors, and (2) truncating the anomaly edges based on contextual affinity.

First, we calculate the cosine similarity between the attribute features $\mathbf{X}$. 
Then, the result is combined with the original adjacent matrix $\mathbf{A}$ through the Hadamard Product ($\odot$) to establish a similarity-based graph adjacent matrix $\mathbf{S}$.

\begin{equation}
    \mathbf{S}= \left( ( \mathbf{X} \cdot \mathbf{X}^{\mathtt{T}})\odot (|\mathbf{X}| \cdot |\mathbf{X}^{\mathtt{T}}|)^{-1} \right) \odot \mathbf{A}
    \label{eq: sim_matrix}
\end{equation}


Then, we select top-n nodes similar to each node as its contextual candidate nodes and set the edges to 1.0. Meanwhile, we set the rest of the edges in $\mathbf{S}$ to zero and create a contextual adjacent matrix $\hat{\mathbf{A}}$, as Eq.(\ref{eq: preserve_adjacent}).


\begin{equation}
\begin{aligned}
\hat{\mathbf{A}}_{ij}=
    \begin{cases}
        1,\quad{{ \mathbf{S}}}_{ij}\in\text{top-n}({ \mathbf{S}}_{i})\\0, \quad{{\mathbf{S}}}_{ij}\notin\text{top-n}({{\mathbf{S}}}_{i})
    \end{cases}, ~ n = \frac{|\mathcal{N}(v_i)|}{2},
\end{aligned}
\label{eq: preserve_adjacent}
\end{equation}

where $\text{top-n}({ \mathbf{S}}_{i})$ refers to $\text{top-n}$ largest similarity in $ \mathbf{S}_i$ and $\mathcal{N}(v_i)$ is the neighbors of node $v_i$. Thus, it is easy to observe that the contextual adjacency matrix $\hat{\mathbf{A}}$ consists of the relation between the nodes and their contextual neighbors. 

After, similar to the similarity-based graph adjacent matrix $S$, we create an Euclidean Distance-based graph adjacent matrix $\mathbf{E}$,  as Eq.(\ref{eq: Distance}).

\begin{equation}
\begin{aligned}
\mathbf{E}=\widehat{\mathbf{E}} \odot \mathbf{A}, ~\widehat{\mathbf{E}}_{i,j} = \sqrt{\sum_{k=1}^{d} (x_{i,k} - x_{j,k})^2}
\label{eq: Distance}
\end{aligned}
\end{equation}
where $x_{i,k}$ represents the values of nodes $v_i$ in the $k$-th feature dimension, and $d$ is the dimensionality of the feature space. 
The given formulation represents the matrix $\mathbf{E}$ is derived by the Hadamard Product of the distance matrices $\widehat{\mathbf{E}}$ and the adjacency matrix $\mathbf{A}$, and the elements of matrix $\widehat{\mathbf{E}}$ are calculated using the Euclidean distance. 

Then, we combine the distance matrix $\mathbf{E}$ with the similarity matrix $\mathbf{S}$ to compute the contextual affinity matrix $\mathbf{C}$, which is formulated as follows:

\begin{equation}
\begin{aligned}
\mathbf{C}= (\mathbf{1}-\mathtt{Normalize}(\mathbf{E})) \odot \mathbf{S}
    \label{eq: ca}
\end{aligned}
\end{equation}

where $\mathtt{Normalize}(\cdot)$ is the normalization operation to transfer the value space of $\mathbf{E}$ the same as $\mathbf{S}$. 
Next, we truncate the anomaly edges based on the contextual affinity matrix $\mathcal{\mathbf{C}}$ and remove anomaly edges more accurately by considering both Euclidean distance and Cosine similarity.

\begin{equation}
\begin{aligned}
    \mathbf{A}_{i,j}^{\circledast} &= \sigma \left( \theta_{ij} \hat{\mathbf{A}}_{i,j}\right)\\
    \theta_{ij} &=  \mathbf{C}_{i,j} - \frac{\sum_j \mathbf{C}_{i,j}}{|\mathcal{N}(v_i)|} 
\end{aligned}
\label{eq: truncation matrix}
\end{equation}




Here, $ \sigma(\cdot) $ is the binary step function where $\sigma(x) = 1 $ if $ x > 0 $, and $ \sigma(x) = 0 $ if $ x \leq 0 $. Besides, $\mathcal{N}(v_i)$ denotes the neighbors of a node $v_i$. Since the term
$\frac{\sum_j \mathbf{C}_{i,j}}{|\mathcal{N}(v_i)|} $ calculates the average affinity of node $v_i$, $\theta_{ij}$ represents the truncation when the affinity between $v_i$ and $v_j$ is lower than the average affinity. Based on the above analysis, the Eq.~(\ref{eq: truncation matrix}) represents the further truncation if an edge is not selected in the contextual adjacent matrix $\hat{\mathbf{A}}$.
Meanwhile, the maximum truncation is limited to the ratio $\leq \beta$ of the total edges of the graph.
Here, $\beta$ is a hyperparameter that will be discussed in the experiment.
The outcome, $\mathbf{A}^{\circledast}$, is a final truncated adjacency matrix that effectively filters out edges failing to meet the specified contextual affinity threshold.


Finally, the proposed contextual affinity truncation mechanism, which integrates contextual affinity with Euclidean distance, enables more precise preservation of normal edges while effectively truncating anomalous edges to reduce the affinities of anomalous nodes. The resulting truncated graph ${G}^{\circledast}$ is defined as follows:

\begin{equation}
G^{\circledast} = \{\mathcal{V}, \mathbf{X}, \mathbf{A}^{\circledast}\}
\label{eq: Truncated graph}
\end{equation}







\subsection{GAT: Global Affinity Truncation }
Based on the assumption that similar nodes are more likely to have normal edge relationships, we designed the global affinity truncation graph.
This graph further enhances the contextual truncation module by increasing the affinities of normal nodes. Two key processes are involved to synthesize this graph effectively: (1) node feature projection, which parametrically projects node features to a suitable space, and (2) global affinity adjacency matrix synthesis, which constructs adjacency matrices based on pairwise similarity.

First, we applied a multilayer perceptron (MLP) to project the features of nodes with different feature distributions $\mathbf{X}$ into a unified feature space $\mathbf{Z}$, which is computed by:

\begin{equation}
{\mathbf{Z}} = \mathtt{MLP}(\mathbf{X})
\label{eq: projection}
\end{equation} 

To construct the global affinity adjacent matrix, we then calculate the node-to-node similarity based on the unified feature space $\mathbf{Z}$. The equation~(\ref{eq: similarity}) presents the process for calculating the node similarity:

\begin{equation}
\bar{\mathbf{S}} = \mathbf{Z} \cdot \mathbf{Z}^{\mathtt{T}}
\label{eq: similarity}
\end{equation}
where $\bar{\mathbf{S}}\in\mathbb{R}^{N \times N}$ is the global affinity matrix and $N$ is the total number of graph nodes.

The global affinity adjacent matrix $\bar{\mathbf{S}}$ is usually dense and represents a fully connected graph, which is often not meaningful for most applications and can lead to expensive computational costs. Therefore, we apply the k-nearest neighbors (kNN)-based sparsification on $\bar{\mathbf{S}}$. Specifically, we retain the edges with the top-k connection values for each node and set the rest to zero. Let ${\mathbf{A}}^{*}$ represent the sparse global affinity matrix, which is defined as:

\begin{equation}
\begin{aligned}
\mathbf{A}^*_{ij}&=
\begin{cases}
1,&\quad{\bar{\mathbf{S}}}_{ij}\in\text{top-k}({\bar{\mathbf{S}}}_{i}),\\0,&\quad{\bar{\mathbf{S}}}_{ij}\notin\text{top-k}({\bar{\mathbf{S}}}_{i}),
\end{cases}
\end{aligned}
\label{eq: adjacent}
\end{equation}
where $\text{top-k}({{\mathbf{S}_i}})$ is the set of top-k values of row vector ${\bar{\mathbf{S}}}_{i}$. 

At last, by combining the graph-independent node feature $\mathbf{Z}$ and the sparse adjacent matrix $\bar{\mathbf{A}}$, we have the global affinity truncation graph $G^{*}$ which can be expressed as follows:

\begin{equation}
G^{*} = \{\mathcal{V}, \mathbf{X}, \mathbf{A}^*\}
\label{eq: global affinity graph}
\end{equation}



\subsection{Affinity Combined Maximization}
So far, we have obtained the contextual truncation graph $G^{\circledast}$ and the global affinity truncation graph $G^{*}$. GCTAM is designed to learn a GNN-based affinity combined maximization model that maximizes the affinity of normal nodes while decreasing the affinity of anomaly nodes. Specifically, the projection from the graph nodes onto new representations using GNN layers. In this work, we employ graph convolution networks (GCNs) \cite{kipf2016gcn} due to their computational efficiency. The node representations are computed as follows:

\begin{equation}
\begin{aligned}
    \mathbf{H}^{\circledast} =& \mathtt{GNN}^{\Theta}(G^{\circledast}) = \left[\begin{matrix}\mathbf{h}^{\circledast}_{1}\\\vdots\\ \mathbf{h}^{\circledast}_{N}\end{matrix}\right] \\
    \mathbf{H}^{*} =& \mathtt{GNN}^{\Theta}(G^{*}) = \left[\begin{matrix}\mathbf{h}^{*}_{1}\\\vdots\\ \mathbf{h}^{*}_{N}\end{matrix}\right] \\
\end{aligned}
\label{eq: gnn}
\end{equation}
where $\Theta$ denotes the shared learnable parameters. Thus, we can notice that $\mathbf{H}^{\circledast}$ and $\mathbf{H}^{*}$ represent the node representations obtained from the contextual truncation graph $G^{\circledast}$ and the global affinity truncation graph $G^{*}$, respectively.

\textbf{Affinity Score Calculation.} We calculate the local affinity of each node to its neighbors to exploit this homophily property for unsupervised GAD. The local affinity can be defined as an averaged affinity to the neighboring nodes:

\begin{equation}
\label{eq: affinity socre}
\begin{aligned}
\mathcal{AS}^{\circledast}({v_i}) =& \frac{1}{{\left| {{\cal N^{\circledast}}\left( {{v_i}} \right)} \right|}} \sum\limits_{{v_j} \in {\cal N^{\circledast}}\left( {{v_i}} \right)} {\frac{{\mathbf{h}_i^{\circledast} \cdot{{{\mathbf{h}^{\circledast}}}_j}}}     { |{\mathbf{h}}^{\circledast}_i |  |{\mathbf{h}^{\circledast}_j}| }},
\\   
\mathcal{AS}^{*}({v_i}) =& \frac{1}{{\left| {{\cal N^{*}}\left( {{v_i}} \right)} \right|}}\sum\limits_{{v_j} \in {\cal N^{*}}\left( {{v_i}} \right)} {\frac{{\mathbf{h}_i^{*} \cdot{{{\mathbf{h}^{*}}}_j}}}     { |{\mathbf{h}}^{*}_i |  |{\mathbf{h}^{*}_j}| }}
\end{aligned}
\end{equation}
where ${\cal N^{\circledast}}({v_i})$ and ${\cal N^{*}}({v_i})$ are the neighbor sets of node $v_i$ in the contextual affinity truncation graph $G^{\circledast}$ and the global affinity truncation graph $G^{*}$, respectively. Similarly, $\mathcal{AS}^{\circledast}({v_i}) $ and $\mathcal{AS}^{*}({v_i})$ represent the affinity score of the node $v_i$ in the two types graphs. The larger the affinity, the more likely the node is normal, while a smaller affinity indicates a higher probability that the node is anomalous.

\textbf{Learning Objective.} Finally, our optimization objective is maximizing the node affinities based on the contextual affinity truncation graph $G^{\circledast}$ and global affinity truncation graph $G^{*}$. Since we reduce the affinities of anomalous nodes in the contextual affinity truncation graph, we enhance the affinities of normal nodes in the global affinity truncation graph. By maximizing the node affinities, we can effectively distinguish between normal and anomalous nodes. The optimization objective function is formulated as follows:

\begin{equation}
\begin{aligned}
    \mathcal{L} &=\min\limits_{\Theta}\left( - \sum_{i=1}^{N} \left(\mathcal{AS}^{\circledast}({v_i}) + \mathcal{AS}^{*}({v_i}) \right) \right)
\end{aligned}
\label{eq: objective}
\end{equation}

By optimizing the above equation, the local affinity of each node can be maximized based on the learned node representations, $\mathbf{H}^{\circledast}$ and $\mathbf{H}^{*}$, derived from the contextual affinity truncation graph $G^{\circledast}$ and the global affinity truncation graph $G^{*}$.

\section{Experiment}
\subsection{Experiment Settings}
\noindent\textbf{Datasets.}
 We conduct the experiments on seven widely used publicly available real-world GAD datasets from diverse social networks, and citation networks, including Amazon~\cite{dou2020enhancing}, YelpChi~\cite{kumar2019predicting}, ACM~\cite{tang2008arnetminer}, Facebook~\cite{xu2022contrastive}, and Reddit. The ACM dataset contains two types of injected anomalies, contextual and structural anomalies ~\cite{Cola_liu2021anomaly,dominant_ding2019deep}, that are nodes with significantly deviated graph structure and node attributes, respectively. The other six datasets contain real anomalies. The statistics of all datasets are shown in Table~\ref{tab: datasets}. 

\begin{table}[ht!]
\setlength{\tabcolsep}{0.65mm}
\centering
\resizebox{\linewidth}{!}{%
\begin{tabular}{lcccccc}
\toprule
Dataset & Type &R/I & Nodes & Edges & Attributes & Anomalies(Rate)  \\
\midrule
Amazon   & Co-review  & R &10244 &175,608  &25 & 693(6.66\%)\\
YelpChi  & Co-review  &R &24,741 &49,315 &32 &1,217(4.91\%)\\
ACM &Citation Networks & I& 16,484 &71,980 &8,337 &597(3.63\%)\\
Facebook  & Social Networks& R& 1,081 &55,104 &576 &27(2.49\%)\\
Reddit  & Social Networks&R &10,984 &168,016 &64 &366(3.33\%)\\
Amazon-all & Co-review &R    & 11,944 &  4,398,392&25 & 821(6.87\%)   \\   YelpChi-all  & Co-review  &R &45,941 &3,846,979 &32 &6,674(14.52\%)\\
\bottomrule
\end{tabular}
}
\caption{The statistics of seven datasets. The R/I represents the datasets with Injected/Real anomalies.}
\label{tab: datasets}
\end{table}

\noindent\textbf{Baselines.}
GCTAM is compared with a total of eight state-of-the-art (SOTA) methods. The three contrastive-based self-supervised methods are CoLA~\cite{Cola_liu2021anomaly}, SL-GAD~\cite{SL_GAD_zheng2021generative}, and HCM-A~\cite{huang2021hop}, while four are reconstruction-based self-supervised methods, namely DOMINANT~\cite{dominant_ding2019deep}, iForest~\cite{liu2012isolation}, ANOMALOUS~\cite{ANOMALOUS_peng2018anomalous} and ComGA~\cite{ComGA_luo2022comga}. One semi-supervised model, GGAD~\cite{semi_qiao2024generative}. Additionally, GCTAM is compared with the affinity-based SOTA method, TAM~\cite{TAM_qiao2024truncated}.


\begin{table*}[ht!]
\centering
\textit{Note}: * represents reproduced result, others are reported in TAM\cite{TAM_qiao2024truncated}.
\resizebox{\textwidth}{!}{%
\begin{tabular}{c|c|cccccccccc}
\hline
\textbf{Metric} & \textbf{Method} & \textbf{Amazon} & \textbf{YelpChi} & \textbf{ACM} & \textbf{Facebook} & \textbf{Reddit} & \textbf{Amazon-all} & \textbf{YelpChi-all} \\ \hline
\multirow{9}{*}{AUROC} 
& iForest (2012)      & 56.21±0.8  & 41.20±4.0  & 51.18±1.8  & 53.82±1.5  & 43.63±2.0  & -          & -          \\
& ANOMALOUS (2018)    & 44.57±0.3  & 49.56±0.3  & 68.56±6.3  & 90.21±0.5  & 53.87±1.2  & -          & -          \\
& DOMINANT (2019)     & 59.96±0.4  & 41.33±1.0  & 85.69±2.0  & 56.77±0.2  & 55.55±1.1  & -          & -          \\
& CoLA (2021)         & 58.98±0.8  & 46.36±0.1  & 82.33±0.1  & 84.34±1.1  & \textbf{60.28±0.7}  & 74.36±1.2          & 50.77±0.9          \\
& SL-GAD (2021)       & 59.37±1.1  & 33.12±3.5  & 84.79±0.5  & 79.36±0.5  & 56.77±0.5  & -          & -          \\
& HCM-A (2022)        & 39.56±1.4  & 45.93±0.5  & 80.60±0.4  & 73.87±3.2  & 45.93±1.1  & -          & -          \\
& ComGA (2022)        & 58.95±0.8  & 43.91±0.0  & 82.21±2.5  & 60.55±0.0  & 54.53±0.3  & -          & -          \\
& TAM (2023)          & {70.64±1.0}  & {56.43±0.7}  & \underline{88.78±2.4}  & \underline{91.44±0.8}  & \underline{60.23±0.4}  & \underline{84.76±0.7}  & \underline{58.18±0.5} \\

& GGAD (2024)*          & \underline{73.12±0.8}  & \underline{62.80±0.9}  & {36.91±2.1}  & {70.66±3.6}  & {60.07±0.8}  &{82.30±1.1}      & {55.34±1.2} \\

& \textbf{GCTAM(ours)}         & \textbf{84.38±1.4}  & \textbf{79.00±0.5}  & \textbf{90.77±1.5}  & \textbf{92.38±0.3}  & {59.21±0.6}  & \textbf{88.89±0.6}  & \textbf{59.57±0.4}  \\

&  $\Delta$ & \textcolor{red}{$\uparrow$11.26} & \textcolor{red}{$\uparrow$16.20} & \textcolor{red}{$\uparrow$1.99} & \textcolor{red}{$\uparrow$0.94} & \textcolor{green}{↓ 1.07} & \textcolor{red}{$\uparrow$4.13} & \textcolor{red}{$\uparrow$1.39}

\\ \hline

\multirow{9}{*}{AUPRC} 
& iForest (2012)      & 13.71±0.2  & 4.09±0.0   & 3.72±0.1   & 3.16±0.3   & 2.69±0.1   & -          & -          \\
& ANOMALOUS (2018)    & 5.58±0.1   & 5.19±0.2   & 6.35±0.6   & 18.98±0.4  & 3.75±0.4   & -          & -          \\
& DOMINANT (2019)     & 14.24±0.2  & 3.95±2.0   & 44.02±3.6  & 3.14±4.1   & 3.56±0.2   & -          & -          \\
& CoLA (2021)         & 6.77±0.1   & 4.48±0.2   & 32.35±1.7  & 21.06±1.7  & \textbf{4.49±0.2}   & 20.12±0.4          & 8.48 ± 0.7         \\
& SL-GAD (2021)       & 6.34±0.5   & 3.50±0.0   & 37.84±1.1  & 13.16±2.0  & 4.06±0.4   & -          & -          \\
& HCM-A (2022)        & 5.27±1.5   & 2.87±1.2   & 34.13±0.4  & 7.13±0.4   & 2.87±0.5   & -          & -          \\
& ComGA (2022)        & 11.53±0.5  & 4.23±0.0   & 28.73±1.2  & 3.54±0.1   & 3.74±0.1   & -          & -          \\
& TAM (2023)          & \underline{26.34±0.8}  & {7.78±0.9}  & \underline{51.24±1.8}  & \underline{22.33±1.6}  & \underline{4.46±0.1}  & \underline{43.46±0.9}  & \underline{18.86±0.6} \\

& GGAD (2024)*          & {19.10±0.8}  & \underline{12.40±1.5}  & {3.00±0.7}  & {6.47±1.1}  & {4.05±0.7}  & {42.19±0.8}  & {17.79±1.5} \\

& \textbf{GCTAM(ours)} & \textbf{50.69±8.1}  & \textbf{16.04±1.0}  & \textbf{52.10±0.7}  & \textbf{22.81±0.6}  & 4.17±0.1  & \textbf{67.18±4.8}  & \textbf{20.13±0.7} \\ 

& $\Delta$ & \textcolor{red}{$\uparrow$24.35} & \textcolor{red}{$\uparrow$3.64} & \textcolor{red}{$ \uparrow$0.86} & \textcolor{red}{$\uparrow$0.48} & \textcolor{green}{↓  0.32} & \textcolor{red}{$\uparrow$23.72} & \textcolor{red}{$\uparrow$1.27} \\
\bottomrule
\end{tabular}%
}

\caption{AUROC and AUPRC results on seven real-world GAD datasets with injected/real anomalies. The best performance per row is boldfaced, with the second-best underlined. $-$ indicates that the result is not available in the TAM. $\Delta$ represents the improvement (\textcolor{red}{$\uparrow$}) or degradation (\textcolor{green}{↓}) compared to the current best baseline method.}
\label{tab: main_result}
\end{table*}

\noindent\textbf{Evaluation Settings.} 
Following~\cite{TAM_qiao2024truncated,chai2022can,pang2021toward,wang2022cross,zhou2022unseen}, two popular and complementary evaluation metrics for anomaly detection, Area Under the Receiver Operating Characteristic Curve (AUROC) and Area Under the precision-recall curve (AUPRC), are used. Higher AUROC/AUPRC indicates better performance. The reported average and standard deviation of AUROC and AUPRC results are averaged over 5 runs with different random seeds. 

\subsection{Main Results}
The AUROC and AUPRC results for the seven real-world GAD datasets are reported in Table \ref{tab: main_result}. For most of the datasets, GCTAM consistently outperforms all competing methods in both AUROC and AUPRC. However, it achieved weaker performance than CoLA and TAM, which may be caused by the low homophily distribution of normal nodes and anomaly nodes in the Reddit dataset.
Moreover, we observe that the TAM model achieves the second-best performance among all compared methods, highlighting that the truncated affinity maximization approach is better suited for node-level graph anomaly detection than other methods.
Notably, on the challenging Amazon dataset, GCTAM achieves a substantial improvement of +11.26 in AUROC and +24.35 in AUPRC over GGAD, the best-performing competitor. Similarly, on the Yelp dataset, GCTAM demonstrates a remarkable enhancement of +16.20 in AUROC and +3.64 in AUPRC over GGAD. Moreover, for the large-scale Amazon-all dataset, GCTAM outperforms TAM by +4.13 in AUROC and +23.72 in AUPRC. 
This result indicates the superior capabilities of integrating global and contextual information of graphs, particularly on challenging and large-scale datasets, underscoring its potential as a robust solution for Affinity-based anomaly detection tasks.

\begin{table}[htbp!]
\centering
\resizebox{\columnwidth}{!}{%
\begin{tabular}{cccccc}
\toprule
\textbf{Datasets}  & w/o $G^{\circledast} \& {G}^{*}$ & w/$G^{\circledast}$  & w/${G}^{*}$   & GCTAM    \\ 
\midrule
Amazon              & 83.86     & 83.90  &\underline{84.07}   & \textbf{84.38} \\ 
\midrule
YelpChi             & 69.73    & {73.07}  &\underline{77.43}  & \textbf{79.00} \\  \midrule
ACM                 & 84.30    & \underline{89.06}    & 88.70  & \textbf{90.77} \\ 
\midrule
Facebook            & 89.03    & \underline{91.55}    & 89.67   & \textbf{92.38} \\ \midrule
Reddit              & 57.09     & \underline{58.95}   & 57.21   & \textbf{59.21} \\ \midrule
Amazon-all          & 87.32     & 88.14   & \underline{88.41}   & \textbf{88.89} \\ 
\midrule
YelpChi-all         & 58.10      & 58.42  & \underline{59.02}  &\textbf{59.57} \\ 
\bottomrule
\end{tabular}%
}
\caption{The evaluation of CAT $G^{\circledast}$ and GAT ${G}^{*}$ module over the metric AUROC.}
\label{tab: ablation_study}

\end{table}

\subsection{Ablation Study}


In ablation experiments, we evaluate the effectiveness of GCTAM with metric AUROC by excluding the CAT and GAT modules, respectively. For clarity, the GCTAM with or without two modules, which is represented by $G^{\circledast}$ and $G^{*}$, is denoted as w and w/o. As shown in Table~\ref{tab: ablation_study}, GCTAM with two modules achieves the best performance across all datasets. It is worth mentioning that removing either the CAT or GAT module reduces the performance of the model to varying degrees. Specifically, on the ACM, Facebook, and Reddit datasets, the use of the contextual truncation graph results in a significant improvement in AUROC. This demonstrates that the proposed contextual affinity truncation effectively reduces anomaly affinity by successfully truncating anomalous edges. Meanwhile, on the Amazon and YelpChi datasets, a significant improvement in AUROC is observed after applying the global affinity truncation. The main reason is that our global affinity truncation enhances the connections between normal nodes, thereby increasing their affinity scores. In summary, these findings demonstrate that the combination of CAT and GAT modules significantly enhances the performance of affinity-based anomaly detection, further boosting the overall performance of the GCTAM.

\subsection{Parameter Analysis}
In this subsection, we explore the sensitivity of two important hyper-parameters in GCTAM: edge truncation ratio $\beta$ and the number of global affinity truncation graph neighbors $k$. The parameter $\beta$ determines the max number of truncated edges ratio from the original graph, and the parameter $k$ determines the number of neighbor nodes to be used in global affinity adjacent matrix $\mathbf{A}^{*}$ based on k-nearest-neighbor topk$(\cdot)$ (kNN).

\begin{figure}[h]
    \centering
    \begin{minipage}{0.98\linewidth}
        \centering
        \begin{minipage}{0.49\linewidth}
            \centering
            \includegraphics[width=\linewidth]{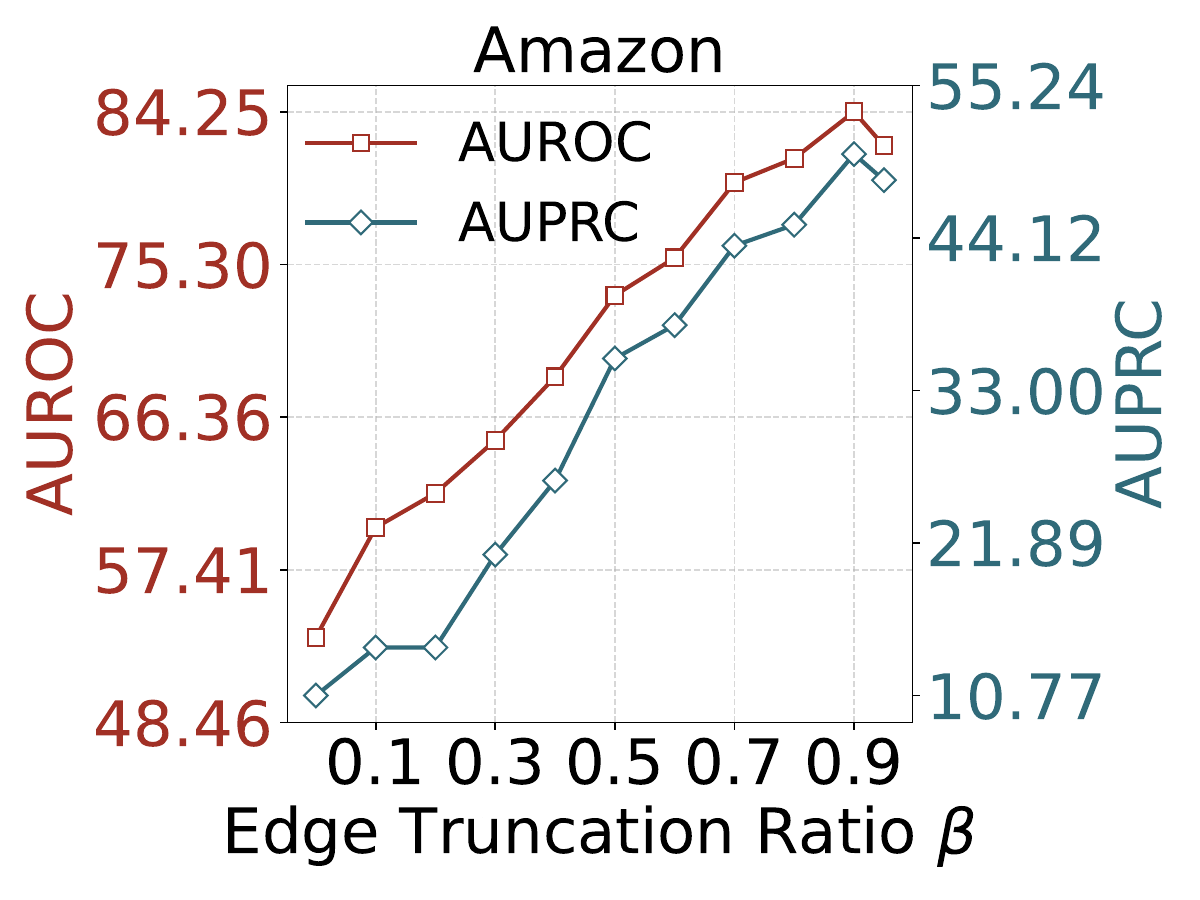}
        \end{minipage}
        \begin{minipage}{0.49\linewidth}
            \centering
            \includegraphics[width=\linewidth]{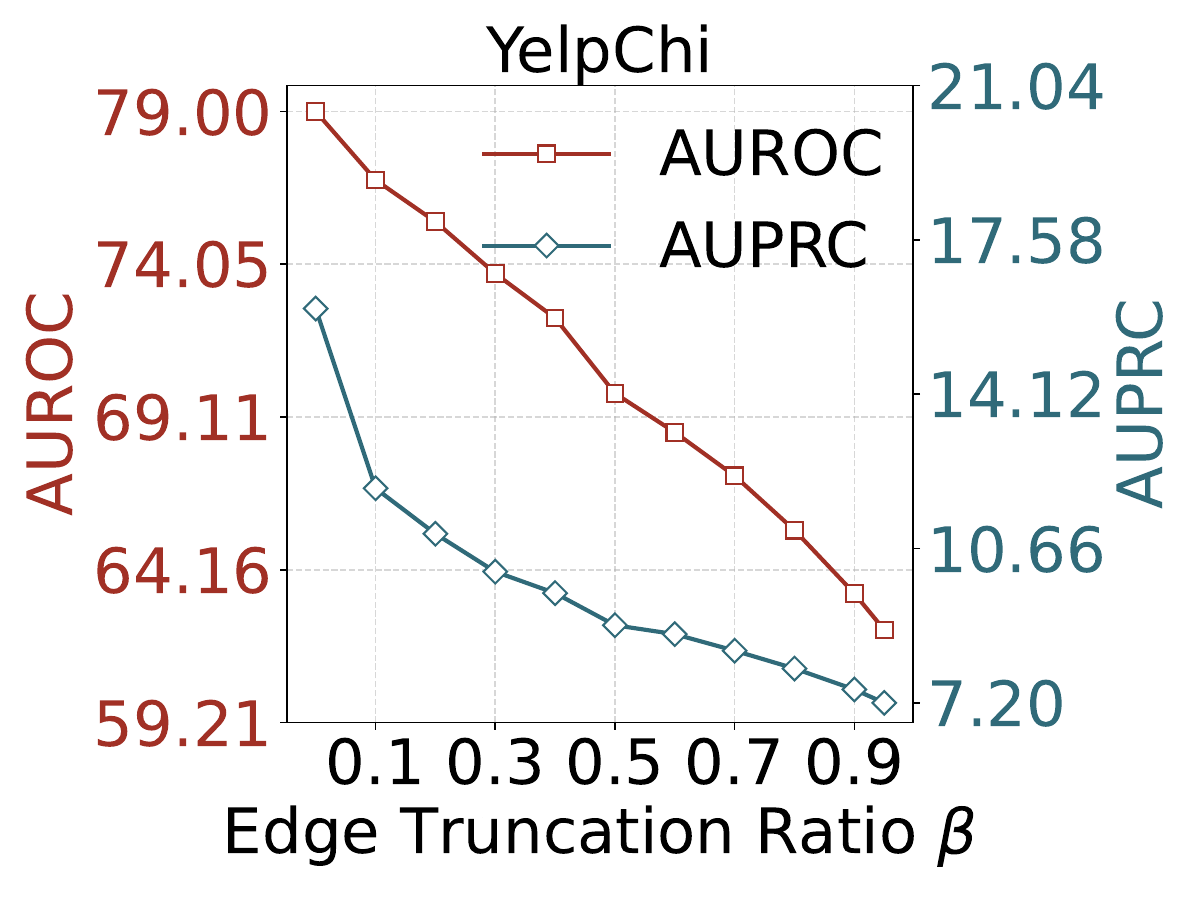}
        \end{minipage}
    \end{minipage}
    \caption{Edge Truncation Ratio VS  AUROC and AUPRC}
    
    \label{fig: edge_truncation_ratio}
\end{figure}

\textbf{Edge truncation ratio $\beta$ analysis.} As shown in the Fig \ref{fig: edge_truncation_ratio}, all datasets were searched for the edge truncation ratio $\beta$, in the range of 0-0.95. From the figure, we can find that different datasets have different sensitivities to the edge truncation ratio $\beta$. Specifically, for the Amazon dataset, the number of anomalous edges is high, and a larger truncation ratio $\beta$ is needed to ensure that the characterization obtained by the model from the truncated graph $G^{\circledast}$ is more conducive to distinguishing anomalous nodes. For the YelpChi dataset, the number of anomalous edges is small, and a better characterization to distinguish anomalous nodes can be obtained based on a smaller edge truncation ratio $\beta$. Therefore, choosing the appropriate edge truncation ratio $\beta$ for different datasets is crucial for affinity-based anomaly node detection.


\begin{figure}[H]
    \centering
    \begin{minipage}{0.99\linewidth}
        \centering
        \begin{minipage}{0.49\linewidth}
            \centering
            \includegraphics[width=\linewidth]{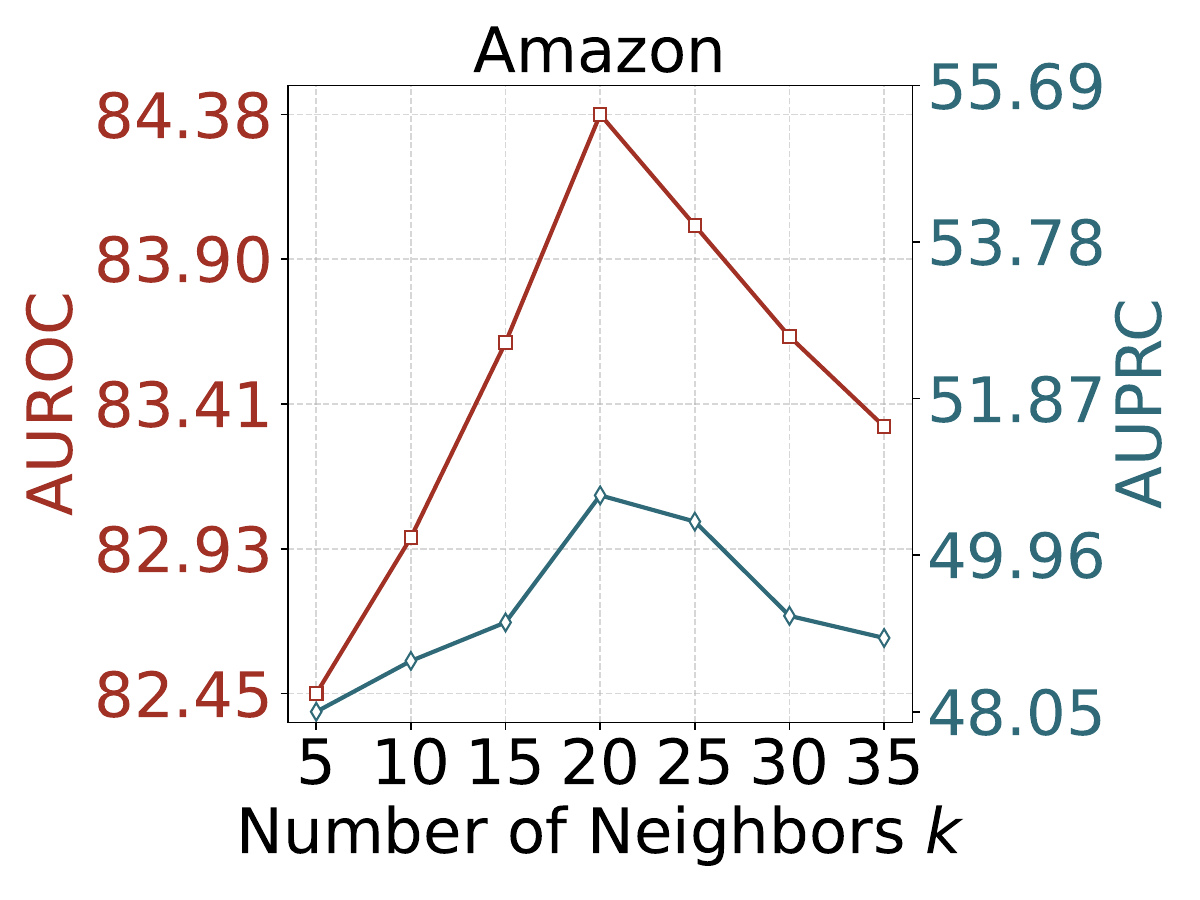}
        \end{minipage}
        \begin{minipage}{0.49\linewidth}
            \centering
            \includegraphics[width=\linewidth]{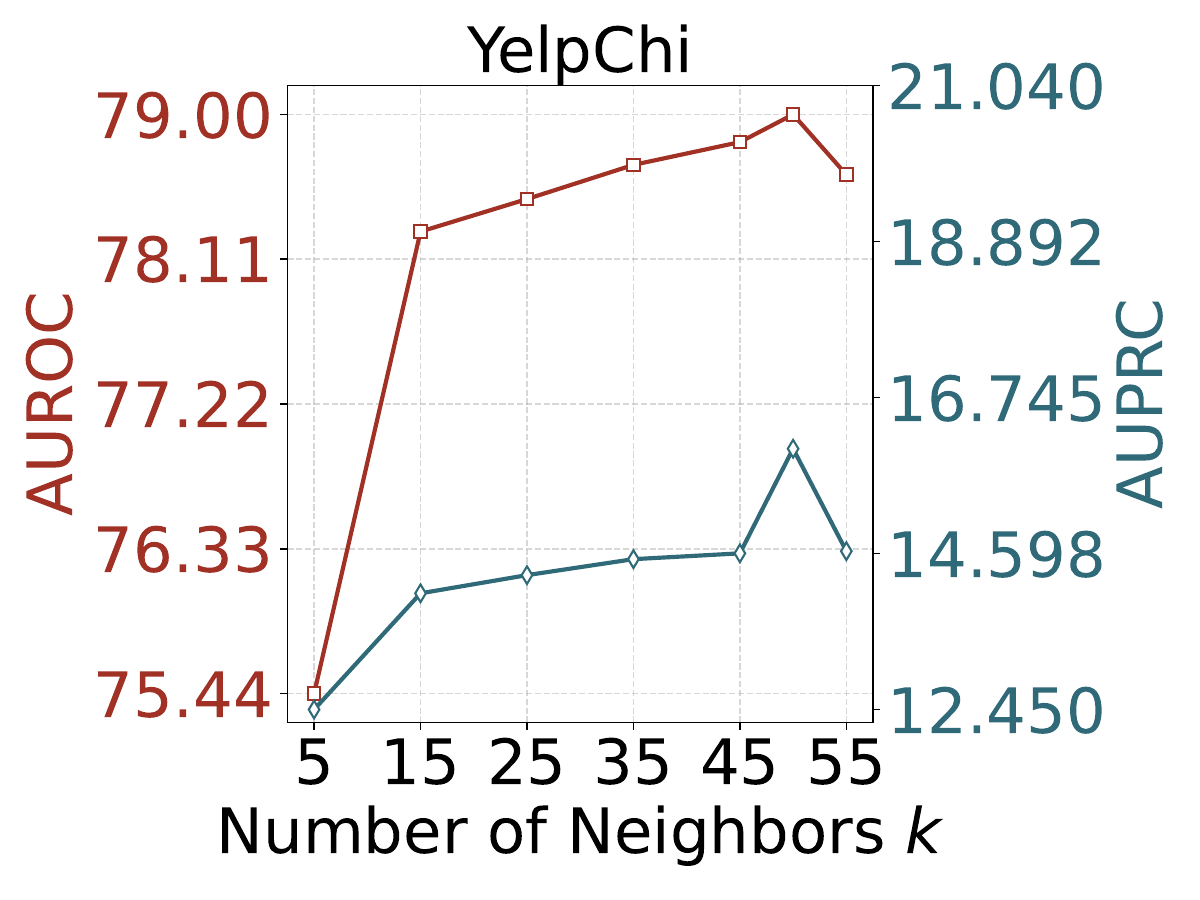}
        \end{minipage}
    \end{minipage}
    \caption{Number of Neighbors k VS  AUROC and AUPRC}
    \label{fig: number_neighbors}
\end{figure}

\textbf{The number of neighbors k analysis.} As shown in Fig.\ref{fig: number_neighbors}, we explore the sensitivity of the hyper-parameter number of neighbors $k$ in GCTAM. The results indicate that selecting an appropriate k value can significantly enhance the AUROC and AUPRC of GCTAM across various datasets. As is demonstrated in Fig.\ref{fig: number_neighbors}, the best selection for each dataset is different, i.e., $k=20$ for Amazon and $k=50$ for YelpChi. It is commonly observed that selecting a value of $k$ that is either too large or too small can lead to suboptimal performance. We hypothesize that an excessively small $k$ may restrict the inclusion of beneficial neighbors, while an overly large $k$ might introduce redundant connections, thereby degrading the overall performance.

\subsection{Affinity Truncation Score Analysis}
To quantitatively assess the quality of node affinity distributions generated by TAM and GCTAM methods, we introduce a novel evaluation metric termed affinity truncation score ($\mathcal{S}_\text{truncation}$). This metric quantifies the discriminative capability of the model by measuring the percentage of normal nodes that maintain higher affinity values compared to anomalous nodes in the learned affinity space. The affinity truncation score is formally defined as follows: 

\begin{equation}
    \mathcal{S}_{\text{truncation}}= \frac{\sum_{i=1}^{|{\mathcal V}_{n}|} \sum_{j=1}^{|{\mathcal V}_{a}|} {1} \quad if( \mathcal{AS}(v_i) > \mathcal{AS}(v_j))}   {|{\mathcal V}_{n}| \cdot |{\mathcal V}_{a}| }
    \label{eq: affinity truncation score}
\end{equation}

where $\mathcal{AS}(\cdot)$ represent the affinity score as defined in Eq.~(\ref{eq: affinity socre}), $|{\mathcal V}_{n}|$ and $|{\mathcal V}_{a}|$ denote the total number of normal and anomalous nodes.  

\begin{figure}[]
    \centering
    \includegraphics[width=1 \linewidth]{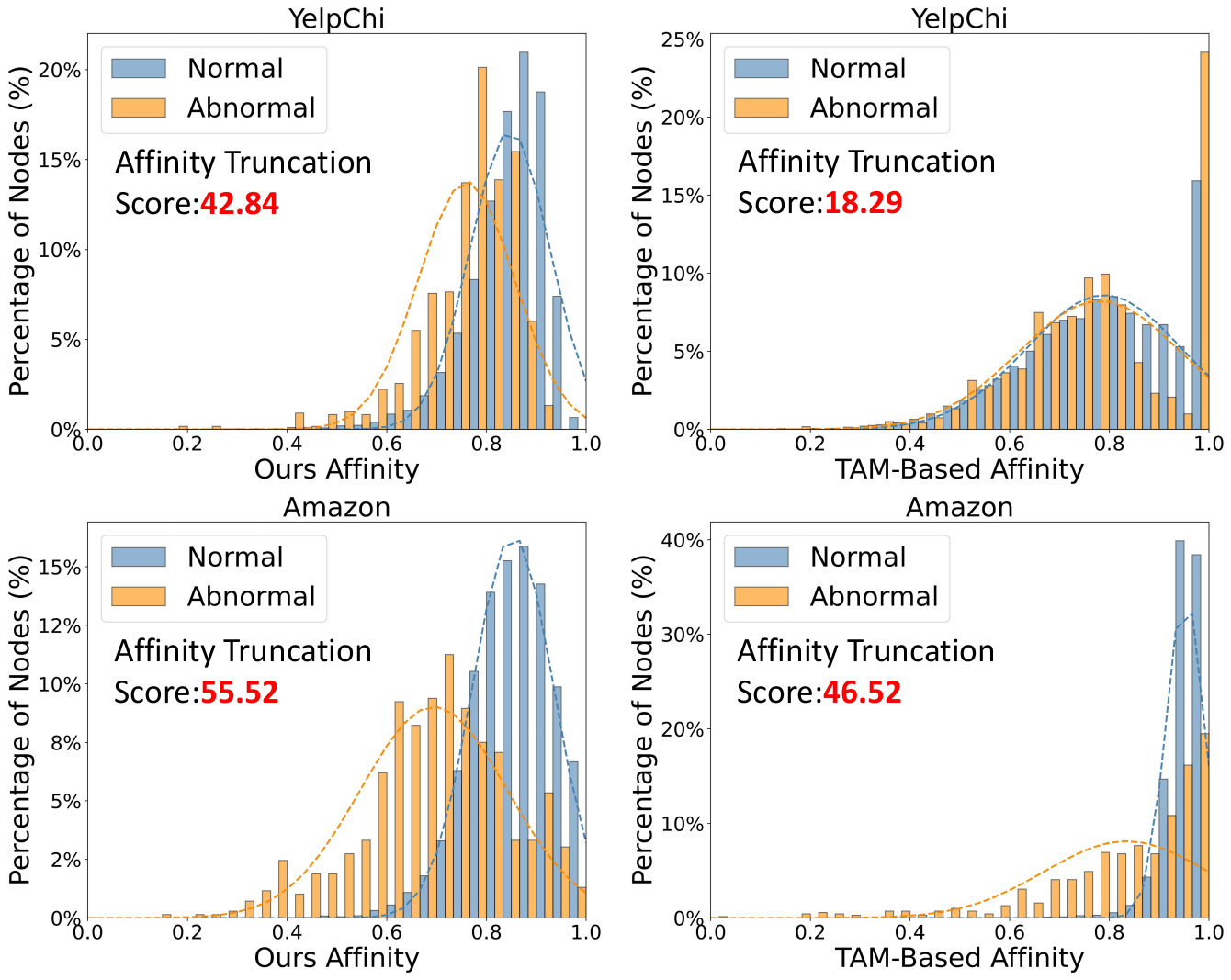}
    \caption{Affinity Truncation Score Compare}
    \label{fig: affinity compare}
\end{figure}

As illustrated in Fig.~\ref{fig: affinity compare}, we conduct a comprehensive comparison of node affinity distributions between normal and anomalous nodes in both TAM and GCTAM representation spaces. The experimental results demonstrate that GCTAM significantly outperforms TAM in achieving separable node affinity distributions between normal and abnormal nodes. Specifically, on the YelpChi dataset, GCTAM achieves a high-affinity truncation score of 42.84, which is much higher than that of TAM of 18.29. Furthermore, on the Amazon dataset, GCTAM attains an affinity truncation score of 55.52, outperforming TAM's score of 46.52 by a margin of 9 points. These findings highlight the effectiveness of GCTAM in constructing a more discriminative affinity space, particularly in scenarios with challenging anomaly detection tasks. This further demonstrates the superiority of the proposed global and contextual truncated affinity combined maximization model.


\section{Conclusion}
In this paper, we propose a novel framework, Global and Contextual Truncated Affinity Maximization (GCTAM), for unsupervised graph anomaly detection, leveraging the homogeneity assumption to optimize node affinity. GCTAM jointly optimizes the proposed affinity score in an end-to-end manner on both the contextual affinity truncation and the global affinity truncation modules. This approach effectively eliminates anomalous edges to reduce anomaly affinity while simultaneously enhancing normal edges to increase the affinity of normal nodes. Extensive experimental evaluations on seven real-world GAD datasets demonstrate that GCTAM outperforms eight competing models. Notably, on challenging datasets Amazon and YelpChi, GCTAM achieves AUROC and AUPRC improvements of +15\% $\sim$ +20\% compared with the previous state-of-the-art methods. The code can be found at \url{https://github.com/kgccc/GCTAM}.

In the future, we will study some adaptations that will work well in some strong heterophily datasets.




\label{sec:Acknowledgments}
\section*{Acknowledgments}
This paper is the result of the research project funded by the National Natural Foundation of China (Grant No. 62106216 and 62162064) and the Open Foundation of Yunnan Key Laboratory of Software Engineering under Grant No.2023SE104.

\label{sec: contribution}
\section*{Contribution Statement}
Xiong Zhang$^{{\dagger}}$ and Hong Peng$^{{\dagger}}$ contributed equally to this work. They were responsible for the conceptualization, methodology, and original draft writing of this study. Cheng Xie$^*$ contributed to the conceptualization, review, and editing of the manuscript and project supervision. The remaining authors provided writing guidance throughout the research.




\bibliographystyle{named}
\bibliography{ijcai25}

@article{twbot22_feng2022twibot,
  title={Twibot-22: Towards graph-based twitter bot detection},
  author={Feng, Shangbin and Tan, Zhaoxuan and Wan, Herun and Wang, Ningnan and Chen, Zilong and Zhang, Binchi and Zheng, Qinghua and Zhang, Wenqian and Lei, Zhenyu and Yang, Shujie and others},
  journal={Advances in Neural Information Processing Systems},
  volume={35},
  pages={35254--35269},
  year={2022}
}

@article{Cola_liu2021anomaly,
  title={Anomaly detection on attributed networks via contrastive self-supervised learning},
  author={Liu, Yixin and Li, Zhao and Pan, Shirui and Gong, Chen and Zhou, Chuan and Karypis, George},
  journal={IEEE transactions on neural networks and learning systems},
  volume={33},
  number={6},
  pages={2378--2392},
  year={2021},
  publisher={IEEE}
}

@inproceedings{dominant_ding2019deep,
  title={Deep anomaly detection on attributed networks},
  author={Ding, Kaize and Li, Jundong and Bhanushali, Rohit and Liu, Huan},
  booktitle={Proceedings of the 2019 SIAM international conference on data mining},
  pages={594--602},
  year={2019},
  organization={SIAM}
}

@article{pourhabibi2020fraud,
  title={Fraud detection: A systematic literature review of graph-based anomaly detection approaches},
  author={Pourhabibi, Tahereh and Ong, Kok-Leong and Kam, Booi H and Boo, Yee Ling},
  journal={Decision Support Systems},
  volume={133},
  pages={113303},
  year={2020},
  publisher={Elsevier}
}

@inproceedings{tang2022rethinking,
  title={Rethinking graph neural networks for anomaly detection},
  author={Tang, Jianheng and Li, Jiajin and Gao, Ziqi and Li, Jia},
  booktitle={International Conference on Machine Learning},
  pages={21076--21089},
  year={2022},
  organization={PMLR}
}

@inproceedings{ComGA_luo2022comga,
  title={Comga: Community-aware attributed graph anomaly detection},
  author={Luo, Xuexiong and Wu, Jia and Beheshti, Amin and Yang, Jian and Zhang, Xiankun and Wang, Yuan and Xue, Shan},
  booktitle={Proceedings of the Fifteenth ACM International Conference on Web Search and Data Mining},
  pages={657--665},
  year={2022}
}

@inproceedings{dae_fan2020anomalydae,
  title={Anomalydae: Dual autoencoder for anomaly detection on attributed networks},
  author={Fan, Haoyi and Zhang, Fengbin and Li, Zuoyong},
  booktitle={ICASSP 2020-2020 IEEE International Conference on Acoustics, Speech and Signal Processing (ICASSP)},
  pages={5685--5689},
  year={2020},
  organization={IEEE}
}

@article{SL_GAD_zheng2021generative,
  title={Generative and contrastive self-supervised learning for graph anomaly detection},
  author={Zheng, Yu and Jin, Ming and Liu, Yixin and Chi, Lianhua and Phan, Khoa T and Chen, Yi-Ping Phoebe},
  journal={IEEE Transactions on Knowledge and Data Engineering},
  volume={35},
  number={12},
  pages={12220--12233},
  year={2021},
  publisher={IEEE}
}

@article{ANEMONE_zheng2021generative,
  title={Generative and contrastive self-supervised learning for graph anomaly detection},
  author={Zheng, Yu and Jin, Ming and Liu, Yixin and Chi, Lianhua and Phan, Khoa T and Chen, Yi-Ping Phoebe},
  journal={IEEE Transactions on Knowledge and Data Engineering},
  volume={35},
  number={12},
  pages={12220--12233},
  year={2021},
  publisher={IEEE}
}

@inproceedings{ANOMALOUS_peng2018anomalous,
  title={ANOMALOUS: A Joint Modeling Approach for Anomaly Detection on Attributed Networks.},
  author={Peng, Zhen and Luo, Minnan and Li, Jundong and Liu, Huan and Zheng, Qinghua and others},
  booktitle={IJCAI},
  volume={18},
  pages={3513--3519},
  year={2018}
}

@article{TAM_qiao2024truncated,
  title={Truncated affinity maximization: One-class homophily modeling for graph anomaly detection},
  author={Qiao, Hezhe and Pang, Guansong},
  journal={Advances in Neural Information Processing Systems},
  volume={36},
  year={2024}
}

@article{chen2022gccad,
  title={GCCAD: Graph Contrastive Learning for Anomaly Detection},
  author={Chen, Bo and Zhang, Jing and Zhang, Xiaokang and Dong, Yuxiao and Song, Jian and Zhang, Peng and Xu, Kaibo and Kharlamov, Evgeny and Tang, Jie},
  journal={IEEE Transactions on Knowledge and Data Engineering},
  year={2022},
  publisher={IEEE}
}

@inproceedings{dou2020enhancing,
  title={Enhancing graph neural network-based fraud detectors against camouflaged fraudsters},
  author={Dou, Yingtong and Liu, Zhiwei and Sun, Li and Deng, Yutong and Peng, Hao and Yu, Philip S},
  booktitle={Proceedings of the 29th ACM International Conference on Information \& Knowledge Management},
  pages={315--324},
  year={2020}
}

@article{zheng2021generative,
  title={Generative and contrastive self-supervised learning for graph anomaly detection},
  author={Zheng, Yu and Jin, Ming and Liu, Yixin and Chi, Lianhua and Phan, Khoa T and Chen, Yi-Ping Phoebe},
  journal={IEEE Transactions on Knowledge and Data Engineering},
  year={2021},
  publisher={IEEE}
}

@inproceedings{duan2023graph,
  title={Graph anomaly detection via multi-scale contrastive learning networks with augmented view},
  author={Duan, Jingcan and Wang, Siwei and Zhang, Pei and Zhu, En and Hu, Jingtao and Jin, Hu and Liu, Yue and Dong, Zhibin},
  booktitle={Proceedings of the AAAI conference on artificial intelligence},
  volume={37},
  number={6},
  pages={7459--7467},
  year={2023}
}

@article{zhang2022reconstruction,
  title={Reconstruction enhanced multi-view contrastive learning for anomaly detection on attributed networks},
  author={Zhang, Jiaqiang and Wang, Senzhang and Chen, Songcan},
  journal={arXiv preprint arXiv:2205.04816},
  year={2022}
}

@inproceedings{tang2008arnetminer,
  title={Arnetminer: extraction and mining of academic social networks},
  author={Tang, Jie and Zhang, Jing and Yao, Limin and Li, Juanzi and Zhang, Li and Su, Zhong},
  booktitle={Proceedings of the 14th ACM SIGKDD international conference on Knowledge discovery and data mining},
  pages={990--998},
  year={2008}
}

@inproceedings{xu2022contrastive,
  title={Contrastive Attributed Network Anomaly Detection with Data Augmentation},
  author={Xu, Zhiming and Huang, Xiao and Zhao, Yue and Dong, Yushun and Li, Jundong},
  booktitle={Pacific-Asia Conference on Knowledge Discovery and Data Mining},
  pages={444--457},
  year={2022},
  organization={Springer}
}

@inproceedings{kumar2019predicting,
  title={Predicting dynamic embedding trajectory in temporal interaction networks},
  author={Kumar, Srijan and Zhang, Xikun and Leskovec, Jure},
  booktitle={Proceedings of the 25th ACM SIGKDD international conference on knowledge discovery \& data mining},
  pages={1269--1278},
  year={2019}
}

@inproceedings{chai2022can,
  title={Can Abnormality be Detected by Graph Neural Networks?},
  author={Chai, Ziwei and You, Siqi and Yang, Yang and Pu, Shiliang and Xu, Jiarong and Cai, Haoyang and Jiang, Weihao},
  booktitle={Proceedings of the Twenty-Ninth International Joint Conference on Artificial Intelligence (IJCAI), Vienna, Austria},
  pages={23--29},
  year={2022}
}

@inproceedings{pang2021toward,
  title={Toward deep supervised anomaly detection: Reinforcement learning from partially labeled anomaly data},
  author={Pang, Guansong and van den Hengel, Anton and Shen, Chunhua and Cao, Longbing},
  booktitle={Proceedings of the 27th ACM SIGKDD conference on knowledge discovery \& data mining},
  pages={1298--1308},
  year={2021}
}

@article{wang2022cross,
  title={Cross-Domain Graph Anomaly Detection via Anomaly-aware Contrastive Alignment},
  author={Wang, Qizhou and Pang, Guansong and Salehi, Mahsa and Buntine, Wray and Leckie, Christopher},
  journal={arXiv preprint arXiv:2212.01096},
  year={2022}
}

@inproceedings{zhou2022unseen,
  title={Unseen Anomaly Detection on Networks via Multi-Hypersphere Learning},
  author={Zhou, Shuang and Huang, Xiao and Liu, Ninghao and Tan, Qiaoyu and Chung, Fu-Lai},
  booktitle={Proceedings of the 2022 SIAM International Conference on Data Mining (SDM)},
  pages={262--270},
  year={2022},
  organization={SIAM}
}

@article{liu2012isolation,
  title={Isolation-based anomaly detection},
  author={Liu, Fei Tony and Ting, Kai Ming and Zhou, Zhi-Hua},
  journal={ACM Transactions on Knowledge Discovery from Data (TKDD)},
  volume={6},
  number={1},
  pages={1--39},
  year={2012},
  publisher={Acm New York, NY, USA}
}

@article{huang2021hop,
  title={Hop-count based self-supervised anomaly detection on attributed networks},
  author={Huang, Tianjin and Pei, Yulong and Menkovski, Vlado and Pechenizkiy, Mykola},
  journal={arXiv preprint arXiv:2104.07917},
  year={2021}
}

@inproceedings{kipf2016gcn,
  title={Semi-supervised classification with graph convolutional networks},
  author={Kipf, Thomas N and Welling, Max},
  booktitle={ICLR},
  year={2016}
}

@inproceedings{zhang2020gcn,
  title={Gcn-based user representation learning for unifying robust recommendation and fraudster detection},
  author={Zhang, Shijie and Yin, Hongzhi and Chen, Tong and Hung, Quoc Viet Nguyen and Huang, Zi and Cui, Lizhen},
  booktitle={Proceedings of the 43rd international ACM SIGIR conference on research and development in information retrieval},
  pages={689--698},
  year={2020}
}

@inproceedings{mukherjee2013yelp,
  title={What yelp fake review filter might be doing?},
  author={Mukherjee, Arjun and Venkataraman, Vivek and Liu, Bing and Glance, Natalie},
  booktitle={Proceedings of the international AAAI conference on web and social media},
  volume={7},
  number={1},
  pages={409--418},
  year={2013}
}

@inproceedings{rayana2015collective,
  title={Collective opinion spam detection: Bridging review networks and metadata},
  author={Rayana, Shebuti and Akoglu, Leman},
  booktitle={Proceedings of the 21th acm sigkdd international conference on knowledge discovery and data mining},
  pages={985--994},
  year={2015}
}

@article{semi_qiao2024generative,
  title={Generative semi-supervised graph anomaly detection},
  author={Qiao, Hezhe and Wen, Qingsong and Li, Xiaoli and Lim, Ee-Peng and Pang, Guansong},
  journal={arXiv preprint arXiv:2402.11887},
  year={2024}
}

@article{44_paszke2019pytorch,
  title={Pytorch: An imperative style, high-performance deep learning library},
  author={Paszke, Adam and Gross, Sam and Massa, Francisco and Lerer, Adam and Bradbury, James and Chanan, Gregory and Killeen, Trevor and Lin, Zeming and Gimelshein, Natalia and Antiga, Luca and others},
  journal={Advances in neural information processing systems},
  volume={32},
  year={2019}
}

@inproceedings{wang2019deep,
  title={Deep graph library: Towards efficient and scalable deep learning on graphs},
  author={Wang, Minjie Yu},
  booktitle={ICLR workshop on representation learning on graphs and manifolds},
  year={2019}
}

\clearpage
\appendix
\begin{appendix}
\section{Affinity Truncation  Score Analysis}
In this section, we calculate an affinity truncation score to demonstrate the effectiveness of GCTAM. Here, RA represents the affinity truncation score derived from the original features and graph structure. Our goal is to enhance the strong affinity between nodes connected by normal edges (normal nodes) while maintaining a weak affinity for nodes connected by anomalous edges (abnormal nodes). A higher affinity truncation score for the truncated graph indicates improved downstream representation and more effective anomaly detection. The affinity truncation score is formally defined in Equation~\ref{eq: affinity truncation score}.        

\begin{table}[H]
\centering
\caption{Comparison of affinity truncation scores.}
\label{tab: affinity truncation scores}
\resizebox{0.8\columnwidth}{!}{%
\begin{tabular}{c|c|c}
\hline
Dataset & Method & Affinity Truncation Score
\\ \hline
\multirow{3}{*}{Amazon} & RA & 26.26 \\
 & TAM & \underline{46.52} \\
 & GCTAM & \textbf{55.52} \\ \hline
\multirow{3}{*}{YelpChi} & RA & \underline{24.75} \\
 & TAM & 18.29 \\
 & GCTAM & \textbf{42.84} \\ \hline
\multirow{3}{*}{Facebook} & RA & 23.80 \\
 & TAM & \underline{62.74} \\
 & GCTAM & \textbf{69.13} \\ \hline
\multirow{3}{*}{Reddit} & RA & 12.20 \\
 & TAM &  \textbf{16.08} \\
 & GCTAM &  \underline{15.52} \\ \hline
\multirow{3}{*}{Amazon-all} & RA & \underline{18.14} \\
 & TAM & 12.57  \\
 & GCTAM & \textbf{66.92} \\ \bottomrule
\end{tabular}%
}
\end{table}

As shown in Table~\ref{tab: affinity truncation scores}, we achieved the highest truncation scores on most of the datasets except Reddit. As discussed previously, this is because Reddit has a low homophily distribution of normal nodes and anomaly nodes dataset.
Specifically, on the Amazon-all dataset, we achieved affinity truncation scores of 66.92, much higher than the 12.57 obtained by TAM and the 18.14 received by the original attribute features.

\section{Description of Datasets}
\label{app: datasets}
Here is a more detailed description of all datasets.
\begin{itemize}
    \item \textbf{Amazon} is a graph dataset capturing the relations between users and product reviews. Following \cite{dou2020enhancing,zhang2020gcn}, three different user-user graph datasets are derived from Amazon using different adjacency matrix construction approaches. 
    In this work, we focus on the Amazon-UPU dataset that connects the users who give reviews to at least one product. The users with less than 20\% are treated as anomalies. 

    \item \textbf{YelpChi} includes hotel and restaurant reviews filtered (spam) and recommended (legitimate) by Yelp. Following \cite{mukherjee2013yelp,rayana2015collective}, three different graph datasets derived from Yelp using different connections in user, product review text, and time.
    
    \item \textbf{ACM} is a citation graph dataset where the nodes denote the published papers and the edge denotes the citation relationship between the papers. The attributes of each node are the content of the corresponding paper.
    
    \item \textbf{Facebook} \cite{xu2022contrastive} is a social network where users build relationships with others and share their friends.

    \item \textbf{Reddit} is a network of forum posts from the social media Reddit, in which the user who has been banned from the platform is annotated as an anomaly. Their post texts were converted to a vector as their attribute.
    
    \item \textbf{Amazon-all and YelpChi-all}  are two datasets by treating the different relations as a single relation following \cite{chen2022gccad}.
\end{itemize}

\section{Implementation Details}
\subsection{Implementation Details.} 
The experiments in this study were conducted on a Linux server running Ubuntu 20.04. The server was equipped with a 13th Gen Intel(R) Core(TM) i9-13900K CPU, 128GB of RAM, and an NVIDIA GeForce RTX 4090 GPU (24GB memory). For software, we used Anaconda3 to manage the Python environment and PyCharm as the development IDE. The specific software versions were Python 3.10.14, CUDA 11.7, DGL 1.01~\cite{wang2019deep} and PyTorch 1.13.1~\cite{44_paszke2019pytorch}. This setup provided a robust and efficient environment for our node classification experiments.

\subsection{Searched Hyperparameters}
\label{app: search hyper}

To facilitate the reproducibility of this work, we list in Table \ref{tab: parameter setting} the GCTAM hyperparameters searched on different datasets. In particular, for the more homophily dataset YelpChi, we conduct a larger exploration of the Number of neighbors in kNN k: \{5,15,25,35,45,50,55,60\}.
The search space is provided as follows:

\begin{table}[ht!]
\centering
\caption{Hyperparameter settings on each dataset.}
\label{tab: parameter setting}
\resizebox{\columnwidth}{!}{%
\begin{tabular}{ccccccc}
\hline
\textbf{Dataset} & Lr & Epochs & Neighbors k & $\beta$ & Layers & HiddenSize \\ \hline
\textbf{Amazon} & 1e-5 & 500 & 20 & 0.9 & 2 & 128 \\
\textbf{YelpChi} & 1e-5 & 500 & 55 & 0.0 & 2 & 128 \\
\textbf{ACM} & 1e-5 & 500 & 25 & 0.3 & 2 & 128 \\
\textbf{Facebook} & 1e-5 & 500 & 20 & 0.3 & 2 & 128 \\
\textbf{Reddit} & 1e-5 & 500 & 20 & 0.1 & 2 & 128 \\
\textbf{Amazon-all} & 1e-5 & 500 & 20 & 0.9 & 2 & 128 \\
\textbf{YelpChi-all} & 1e-5 & 500 & 20 & 0.0 & 2 & 128 \\ \hline
\end{tabular}%
}
\end{table}

\begin{itemize}
    \item Learning rate lr: \{0.01, 0.001, 0.0001 ,0.00001\}
    \item Weight decay : \{1e-3, 1e-4, 1e-5, 1e-6\}
    \item Number of epochs: \{300, 400, 500\}
    \item Number of neighbors in kNN $k$: \{5, 10, 15, 20, 25, 30, 35\}
    \item Edge truncation ratio $\beta$: \{0.0, 0.1, 0.2, 0.3, 0.4, 0.5, 0.6, 0.7, 0.8, 0.9, 0.95\}
\end{itemize}

\subsection{Algorithm}

The training algorithm of \textbf{GCTAM} is summarized in Algorithm \ref{alg:alg1}.
\begin{algorithm}[h!]
\renewcommand{\algorithmicrequire}{\textbf{Input:}}
\renewcommand{\algorithmicensure}{\textbf{Parameters:}}
\caption{The training algorithm of GCTAM}
\label{alg:alg1}

\begin{algorithmic}
\REQUIRE Original graph $G=(\mathcal{V}, A, {X})$;  Feature matrix  ${X}$; Adjacency matrix ${A}$.
\ENSURE Number of epoch $T$ , Edge truncated rate $\beta$;
\end{algorithmic} 

\begin{algorithmic}[1]
\STATE Generate initial parameters for all learnable parameters. 
\WHILE{$\beta < \text{Edge Truncation Threshold}$}
    \STATE Execute contextual affinity truncation (CAT) to obtain graph $G^{\circledast}$ by Eq.(\ref{eq: sim_matrix}) - Eq.(\ref{eq: Truncated graph})
    \STATE Update edges truncation ratio $\beta$ based on current edge state
\ENDWHILE

\FOR{\text{\rm each epoch} $i=0,1,2,...,T$}
\STATE Execute global affinity truncation (GAT)  to obtain graph $G^{*}$ by Eq. (\ref{eq: projection}) - Eq.(\ref{eq: global affinity graph})\\
\STATE Embedding node representations ${H}^{\circledast}$ and ${H}^{*}$ with GCNs by Eq.(\ref{eq: gnn})
\STATE Calculate nodes affinity score by Eq. (\ref{eq: affinity socre})  \\
\STATE Maximize node affinity by Eq.(\ref{eq: objective})
\STATE Update model parameters via gradient descent
\ENDFOR
\end{algorithmic}
\end{algorithm}

\begin{figure*}[ht!]
    \centering
    \includegraphics[width=1\textwidth]{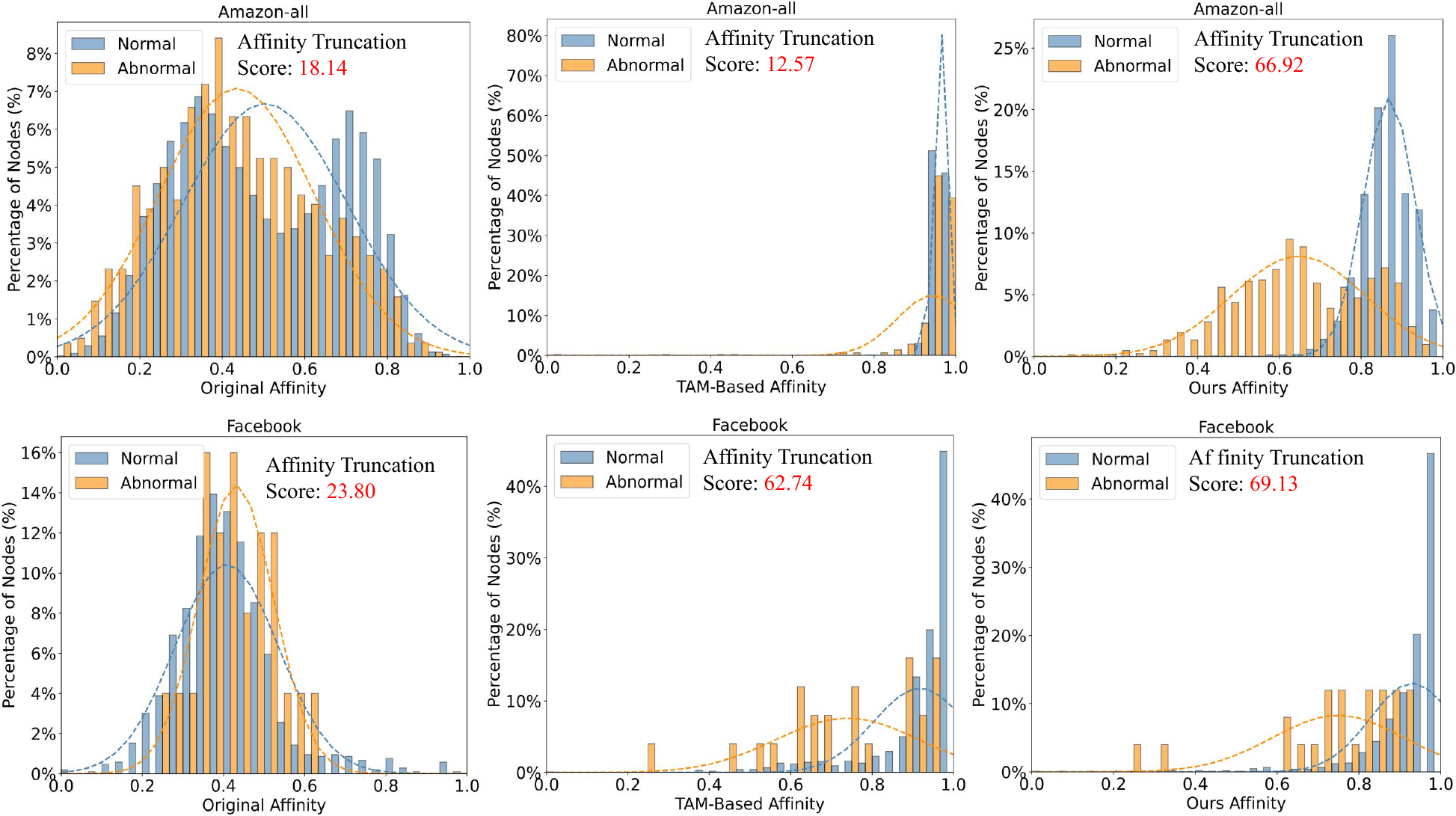}
    \caption{Amazon-all and Facebook  Affinity Truncation Score Compare}
    \label{fig: affinity compare add}
\end{figure*}

\section{Additional Experimental Results}
\subsection{Computational Efficiency}
\noindent\textbf{Computational Efficiency}. The runtime (including both training and inference time) results are shown in Table~\ref{tab: run_time}.  As shown in the results, the training and inference time of our method on all five datasets is much lower than that of the previous SOTA method. Specifically, GCTAM uses only two layers of GCN with shared weights, compared with the TAM method, which needs to perform multiple graph truncations and train multiple LAMNets, so the training and inference time is much less.  Contrastive-based Self-Supervised CoLA method, their time consumption is much higher than that of our method, GCTAM, this is because of the need to construct comparative learning objectives. 

\begin{table}[ht!]
\centering
\caption{Runtime (in seconds) results.}
\label{tab: run_time}
\resizebox{\columnwidth}{!}{%
\begin{tabular}{l|ccccc}
\hline
\multirow{2}{*}{{\textbf{Method}}} & \multicolumn{5}{c}{\textbf{Dataset}} \\
 & Amazon & YelpChi & ACM & Facebook & Reddit \\ \hline
CoLA & 821 & 15340 & 1604 & 65 & 2761 \\
TAM & 142 & 683 & 827 & 23 & 161 \\
GCTAM (Ours) & 86 & 369  & 192  & 5 & 89 \\ \hline
\end{tabular}%
}
\end{table}

\subsection{Affinity Truncation Score Visualization}
As shown in Fig.~\ref{fig: affinity compare add}. We also visualize the affinity truncation scores on Facebook and Amazon-all. The results show that our method achieves very distinguishable affinity truncation scores compared with other methods. In particular, on the Amazon-all dataset, our method achieves a visualization result and affinity truncation score of 66.95, which is much better than that of the current TAM method, which is 12.57.
This can be attributed to the effectiveness of our proposed contextual affinity truncation and global affinity truncation mechanisms, which reduce the affinity of anomalous nodes while enhancing the affinity of normal nodes, thereby improving the discriminative power of the affinity score.

\end{appendix}

\end{document}